\documentclass[preprint,aps,prd,nofootinbib,superscriptaddress]{revtex4}

\usepackage[pdftex]{color,graphicx}
\usepackage{mathrsfs}
\usepackage{epsfig}
\usepackage{amssymb}
\usepackage{amsmath}
\usepackage{feynmp}
\usepackage{slashed}
\usepackage{latexsym}
\usepackage{graphicx}
\usepackage{verbatim}
\usepackage{graphics}
\newcommand{\be}{\begin{equation}}
\newcommand{\ee}{\end{equation}}
\newcommand{\bea}{\begin{eqnarray}}
\newcommand{\eea}{\end{eqnarray}}

\newcommand{\ba}{\begin{array}}
\newcommand{\ea}{\end{array}}

\allowdisplaybreaks
\begin{document}

\title{Can heavy neutrinos dominate neutrinoless double beta decay?}

\author{J. Lopez-Pavon}
\email[]{jlpavon@sissa.it}
\affiliation{IPPP, Department of Physics, Durham University, South Road, Durham, DH1 3LE, UK}
\affiliation{SISSA/ISAS and INFN, 34136 Trieste, Italy}

\author{S. Pascoli}
\email[]{Silvia.pascoli@durham.ac.uk}
\affiliation{IPPP, Department of Physics, Durham University, South Road, Durham, DH1 3LE, UK}

\author{Chan-fai Wong}
\email[]{chan-fai.wong@durham.ac.uk}
\affiliation{IPPP, Department of Physics, Durham University, South Road, Durham, DH1 3LE, UK}

\begin{abstract}
We study whether a dominant contribution to neutrinoless double beta decay coming from extra heavy degrees of freedom, introduced 
to generate the light neutrino masses, can dominate over the light neutrino contribution. It has been shown that this may occur at 
tree-level if the light neutrino contribution partially cancels out. Here we focus on this case, specifically in the context of 
type-I seesaw models paying special attention to the one-loop corrections to light neutrino masses, their contribution to the 
process and correlation with the heavy sector. We perform a general analysis without restricting the study to any particular
region of the parameter space, although interesting limits associated with inverse and extended seesawlike models are discussed 
in more detail. It turns out that the heavy neutrinos can dominate the process only in those limits. For the inverse seesaw limit, 
we find a very constrained allowed region of the parameter space, with heavy neutrino masses around $5$ GeV. The extended 
seesaw case allows for a larger region, but in general, a hierarchical spectrum of heavy neutrinos with masses above and 
below $\sim100$ MeV is required.
\end{abstract}


\preprint{IPPP/12/68}
\preprint{DCPT/12/136}

\maketitle

\section{Introduction}

The existence of neutrino masses, strongly supported by neutrino oscillation experiments, 
is the first experimental evidence of physics beyond the 
Standard Model (SM). 
Furthermore, the fact that neutrino masses are smaller than the masses of the other SM fermions by several orders
of magnitude calls for a ``natural'' 
New Physics (NP) explanation. 
Most of the models, including the very popular seesaw~\cite{Minkowski:1977sc,GellMann:1980vs,Yanagida:1979as,Mohapatra:1979ia} ones, 
assume that the lepton number is not a conserved symmetry and that light neutrinos are Majorana particles.
An interesting experimental window to search for NP gets opened: 
lepton-number-violating processes, highly suppressed in the SM,
among which neutrinoless double beta decay ($0\nu\beta\beta$ decay) experiments are the most promising.
In combination with the information coming from neutrino oscillation experiments, 
absolute neutrino mass experiments, precision measurements, and cosmology, $0\nu\beta\beta$ decay experiments can give us
precious clues in order to identify the mechanism responsible for the neutrino mass generation
and provide a complementary way to look for NP, possibly not otherwise accessible at the LHC.

Although NP is necessary in order to have $0\nu\beta\beta$ decay, 
its effects are usually indirect since the light neutrinos generically dominate 
the process in most of the models, as is the case in type-I, type-II~\cite{Magg:1980ut,Schechter:1980gr,
Wetterich:1981bx,Lazarides:1980nt,Mohapatra:1980yp} and type-III~\cite{Foot:1988aq} seesaw realizations 
and some extradimensional models~\cite{Dienes:1998sb,ArkaniHamed:1998vp,Blennow:2010zu}. The key point is that the 
light neutrino contribution 
and the NP one are usually correlated through the generation of the light neutrino masses and
the second of these, suppressed by being short range, is thus constrained and generally subdominant~\cite{Blennow:2010th}. The 
question of whether a measurable direct contribution 
to the $0\nu\beta\beta$ decay rate coming from NP is theoretically and phenomenologically viable is thus very interesting. 
This question has been addressed recently in Refs.~\cite{Ibarra:2011xn,Mitra:2011qr} in the context of different
type-I seesaw models. In these publications, a relevant exception to the argument above has been pointed out:
the case in which the tree-level light neutrino contribution, 
 induced by the presence of heavy fermion singlets, partially cancels out. In this case, it is found that the direct heavy neutrino contribution
to the process is indeed relevant and can be as large as current bounds. However, no detailed discussion about the 
correlation among the light and heavy contributions, once the one-loop corrections to neutrino masses are included in 
the analysis, is given. 
The main goal of this work is to analyze to what extent having this dominant contribution 
from heavy neutrinos is possible in the general framework of type-I seesaw models 
when the relevant one-loop corrections and experimental constraints are carefully considered, 
paying special attention to its correlation with the light neutrino contribution induced by these
corrections. 

We will first very briefly review 
the aspects of the $0\nu\beta\beta$ decay phenomenology 
relevant for our analysis. Considering a general parameterization 
of the neutrino mass matrix without restricting the analysis to any region of the parameter space,
we will then study under which 
conditions the light neutrino contribution can be canceled at tree-level. We will include the one-loop corrections and study if the heavy 
neutrinos can give a dominant and measurable (i.e. within reach of the next-to-next $0\nu\beta\beta$ decay experiments) contribution to the process. Finally, we will show that, even when the tree-level
cancellation takes place, the light and heavy contributions are not completely decoupled once the one-loop corrections are included 
in the study and a dominant heavy contribution may occur only in specific regions of the parameter space.   
   
This paper is organized as follows: In Sec. II, we briefly review the $0\nu\beta\beta$ decay phenomenology in the general 
context of seesaw models, introducing the notation and the Nuclear Matrix Elements (NMEs) we will use. 
In Sec. III, the parameterization of the neutrino mass matrix is presented, distinguishing some relevant limits and their relation with 
well-known models, such as the inverse and extended seesaw ones. Section IV is devoted to the study of the cancellation condition of the light 
neutrino contribution and its tree-level consequences on the heavy neutrino sector. Section V is dedicated to the analysis of 
the relevant corrections: higher-order corrections to the seesaw expansion and one-loop corrections. The combined analysis of 
the $0\nu\beta\beta$ decay phenomenology, when these corrections and the relevant experimental constraints are taken into account, is presented 
in Sec. V. Finally, in Sec. VI we draw our conclusions.

\section{Dominant heavy neutrino contribution to $0\nu\beta\beta$ decay?}

As we have already mentioned, in the context of seesaw models, the contributions to neutrinoless double beta decay 
from NP at scales much heavier than the exchanged momentum ($\sim100$ MeV), namely the ones mediated
by heavy fermion singlets or scalar/fermion triplets introduced to generate the light neutrino masses,
is usually subdominant and the light neutrinos 
typically dominate the process~\cite{Blennow:2010th}.

Let us very briefly review how the above result is obtained and the possible exceptions. Following the notation in 
Ref.~\cite{Blennow:2010th}, and restricting the study to type-I seesaw models, the $0\nu\beta\beta$ decay rate can be written as
\begin{equation}
\frac{\Gamma_{0\nu\beta\beta}}{\ln2}=
G_{01}\left|\sum_{j}U_{ej}^{2}\frac{m_j}{m_e}\mathcal{M}^{0\nu\beta\beta}(m_j)\right|^{2},
\label{decayrate}
\end{equation}
where $G_{01}$ is a well-known kinematic factor, $U$ is the unitary matrix which diagonalizes the complete neutrino mass matrix both for 
active and sterile neutrinos, $m_{j}$ 
are the corresponding eigenvalues, i.e., the neutrino masses, and $\mathcal{M}^{0\nu\beta\beta}$ are the Nuclear Matrix 
Elements (NMEs) associated with the process. The sum should be made over all the neutrino masses, including the heavy ones.

The NMEs can be computed using different methods, the main two being the quasiparticle random phase approximation (QRPA) 
\cite{Suhonen:1998ck,Faessler:1999zg} and the interacting shell model (ISM) ~\cite{Caurier:2004gf,Caurier:2007wq}. In this work we will 
make use of the NME data presented in Ref.~\cite{Blennow:2010th} and available in Ref.~\cite{nme}. They were computed for different nuclei in the 
context of the ISM as a function of the neutrino mass, something very convenient for our analysis. We use a notation
in which the dependence on the neutrino propagator is included on $\mathcal{M}^{0\nu\beta\beta}(m_j)$, in contrast with the 
notation usually adopted in the literature where the propagator is expanded to factorize the mass dependence.
In Fig. 1 of Ref.~\cite{Blennow:2010th} the NME dependence on the mass of the neutrino mediating the process is depicted, 
showing two different regions separated by the scale of the process $\sim100$ MeV:

\begin{itemize}

\item  Below the $0\nu\beta\beta$ scale, the NMEs reach their maximum value and are mainly independent of the neutrino mass. 
For $m_i\ll 100$ MeV, $\mathcal{M}^{0\nu\beta\beta}(m_i)=\mathcal{M}^{0\nu\beta\beta}(0)$.

\item  The NMEs corresponding to neutrinos much heavier than $100$~MeV are suppressed with the heavy neutrino masses
and scale as $\mathcal{M}^{0\nu\beta\beta}(m_I)\propto 1/m_I^2$. 

\end{itemize}

This behavior of the NMEs, showing two clearly different 
regimes, can be easily understood by expanding the propagator of the neutrino mediating the process. 
The transition region around $100$ MeV is well described in Fig.~1 of Ref.~\cite{Blennow:2010th} since no assumptions 
have been made on the neutrino masses in the NME computation. 

We can distinguish the following two contributions to the $0\nu\beta\beta$ decay amplitude:

\begin{equation}
 A \propto \sum_{i=1}^{3} m_i U_{ei}^2 \mathcal{M}^{0\nu\beta\beta}(m_i) + \sum_I^{\rm extra} m_I U_{eI}^2 \mathcal{M}^{0\nu\beta\beta}(m_I)\,,
\label{typeIamp}
\end{equation}
the first term corresponding to the mostly active neutrino contribution, and the second to the extra 
states of the model. Here and throughout the text, we use capital letters to denote the mass indices of 
the mostly sterile states and lowercase letters for those of the mostly active states.

On the other hand, since a Majorana mass coupling for the active neutrinos is forbidden by the gauge symmetry, the 
diagonalization of the complete mass matrix leads to the following relation:
\begin{equation}
\sum_{i=1}^3 m_i U_{ei}^2 + \sum_I^{\rm extra} m_I U_{eI}^2 = 0.
\label{eq:GIMII}
\end{equation}
%
This equation, which relates the light and extra degrees of freedom of the model, should always be fulfilled at tree-level and
plays a fundamental role in the phenomenology of $0\nu\beta\beta$ decay.

For extra states with all the masses well above $100$ MeV, using the relation given in Eq.~(\ref{eq:GIMII}), 
the contribution to $0\nu\beta\beta$ decay in Eq.~(\ref{typeIamp}) can be recast as

\begin{eqnarray}
 A &\propto& -\sum_I^{\rm heavy} m_I U_{eI}^2 \left(\mathcal{M}^{0\nu\beta\beta}(0) - \mathcal{M}^{0\nu\beta\beta}(m_I)  \right)
\nonumber\\
&\approx& -\sum_I^{\rm heavy} m_I U_{eI}^2 \mathcal{M}^{0\nu\beta\beta}(0) 
= \sum_{i=1}^{3} m_i U_{ei}^2 \mathcal{M}^{0\nu\beta\beta}(0) ~,
\label{eq:heavyfromlight}
\end{eqnarray}
where we have used the fact that $\mathcal{M}^{0\nu\beta\beta}(0) \gg \mathcal{M}^{0\nu\beta\beta}(m_I)$.
The contribution from the light active neutrinos thus dominates. A similar argument applies to models 
which implement the type-II and type-III seesaw~\cite{Blennow:2010th}, and more generically to models in
which heavy sterile neutrino mixing with $\nu_e$ is introduced. As sterile neutrinos contribute to light 
neutrino masses, $\sum_I m_I U_{eI}^2$ is constrained by the value of the light neutrino masses while their 
contribution to $0\nu\beta\beta$ decay is suppressed by 
$\mathcal{M}^{0\nu\beta\beta}(m_I)$ making it subdominant, at least if fine-tuning is not invoked as
we will see in the following.

These considerations apply generically to models with extra sterile neutrinos but there are some notable 
exceptions:

\begin{itemize}
 \item 
The case of extra states below and above $100$ MeV. In this case Eq.~(\ref{typeIamp}) can be rewritten as
\bea
A &\propto& \left(\sum_{i=1}^{3} m_i U_{ei}^2 + \sum_I^{\rm light} m_I U_{eI}^2  \right)
\mathcal{M}^{0\nu\beta\beta}(0)+ \sum_I^{\rm heavy} m_I U_{eI}^2\,\mathcal{M}^{0\nu\beta\beta}(m_I)
\nonumber\\
&\approx& \left(\sum_{i=1}^{3} m_i U_{ei}^2 + \sum_I^{\rm light} m_I U_{eI}^2 \right)\mathcal{M}^{0\nu\beta\beta}(0),\,
\eea
and the new states below $100$ MeV may give the dominant contribution. Notice that if all 
the extra states are below the $0\nu\beta\beta$ scale the cancellation driven by Eq.~(\ref{eq:GIMII}) forbids 
the process. The same behavior as in this type-I seesaw realization with sterile neutrinos below and above the 
$0\nu\beta\beta$ scale applies to a type-II or type-III scenario in combination with type-I
light sterile neutrinos~\cite{Blennow:2010th}. In all these scenarios NP above the $0\nu\beta\beta$ scale, either heavy sterile neutrinos 
(in the type-I seesaw) or heavy triplets (in the just-mentioned mixed type-I/type-II and type-I/type-III seesaw), 
are needed to avoid the cancellation, while the ``light'' sterile neutrinos can give the dominant contribution to $0\nu\beta\beta$
decay.\footnote{The scalar/fermion triplet contribution to $0\nu\beta\beta$ decay is subdominant in comparison
with the light active neutrino one~\cite{Blennow:2010th}.} 

\item 
Additional contributions to neutrino masses. In this case the mass relation becomes
\begin{equation}
\sum_{i=1}^{3} m_i U_{ei}^2 + \sum_I^{\rm heavy} m_I U_{eI}^2 = m_{LL},
\label{eq:GIMIImodified}
\end{equation}
where $m_{LL}$ is an effective Majorana mass term generated for the active neutrinos 
by some other mechanism. $m_{LL}$ and $\sum_I^{\rm heavy} m_I U_{eI}^2$ could be very large 
and cancel nearly exactly, keeping light neutrino masses under control. In this way, even with the contribution to $0\nu\beta\beta$ 
of the heavy states being weighted by the corresponding NME, a dominant effect could arise. However, it would 
have to overcome the suppression coming from the NME ($\mathcal{M}^{0\nu\beta\beta}(0)/\mathcal{M}^{0\nu\beta\beta}(m_I)\gg 1$)
and a very high level of cancellation among $\sum_I^{\rm heavy} m_I U_{eI}^2$ and $m_{LL}$ in Eq.~(\ref{eq:GIMIImodified})
would be required. This possibly implies an uncomfortably high level of fine-tuning and will not be studied in this work. 

\item A cancellation in the light neutrino contribution: $\sum_{i=1}^3 m_i U_{ei}^2 = 0$. 
If this cancellation took place, the heavy neutrinos would trivially dominate the process (at least, at tree-level).

\end{itemize}

In this work we are going to focus on this last possibility. 
This relevant exception was studied in Refs.~\cite{Ibarra:2010xw,Ibarra:2011xn,Mitra:2011qr} and not contemplated
in Ref.~\cite{Blennow:2010th}. Of course, this cancellation in the light contribution could be obtained by invoking some symmetry, 
and the most natural one in this context is the lepton number. The well-known \textit{inverse}~\cite{Mohapatra:1986bd} or 
\textit{linear}~\cite{Malinsky:2005bi} seesaw models, which involve 
small violations of the lepton number, could in principle implement this scenario. 
However, generating a measurable heavy neutrino contribution to the $0\nu\beta\beta$ decay is not trivial even in these models. First of all, 
$0\nu\beta\beta$ decay is a lepton-number-violating process and consequently is expected to be suppressed in this context. Moreover, 
the suppression of the heavy neutrino contribution with the NME ($\sim1/m_I^2$) makes having very low-scale heavy masses unavoidable in order 
to obtain a relevant effect. This possibility has been recently explored claiming that the heavy neutrinos could be very relevant 
for some particular neutrino mass textures~\cite{Ibarra:2011xn,Mitra:2011qr}. Indeed, the main goal of this note is to check 
to what extent this is possible, paying special attention to the stability of the light neutrino masses under one-loop corrections 
and their contribution to the $0\nu\beta\beta$ decay.

\section{The Models}
\label{sec:model}

We will focus on the study of SM extensions which consist of the addition of $n+n'$ fermion gauge singlets, $N_i$, to the 
SM particle content without imposing lepton number conservation, whose Lagrangian is

\begin{eqnarray}
\mathscr{L} &=& \mathscr{L}_\mathrm{SM}+ \mathscr{L}_\mathrm{kin}-\frac{1}{2} \overline{N_{i}} M_{ij} N_{j}^{c}
 -y_{i\alpha}\overline{N_{i}} 
\widetilde \phi^\dagger
L_\alpha +\text{h.c}.,
\end{eqnarray}
where $\mathscr{L}_\mathrm{SM}$ is the SM Lagrangian and $\mathscr{L}_\mathrm{kin}$ are the kinetic 
terms of the new fields $N_i$. Here, and in the rest of the paper, the subindex $\alpha$ denotes 
flavor ($\alpha=e,\mu,\tau$). Without loss of 
generality, the neutrino mass matrix can be expressed as

\begin{eqnarray}
 M_\nu = \left( \begin{array}{ccc} 0 &  Y_1^T v/\sqrt{2}& \epsilon {Y}_2^T v/\sqrt{2} \\  Y_1 v/\sqrt{2}  
&  \mu' & \Lambda \\ \epsilon {Y}_2 v/\sqrt{2}
 & \Lambda^T & \mu  
 \end{array} \right)\equiv \left( \begin{array}{cc} 0 &  m_D^T \\ m_D  
& M 
 \end{array} \right) . 
 \label{Mnu}
\end{eqnarray}
Here $Y_1$ and $Y_2$ are the $n'\times 3$ and $n\times 3$ matrices that form the Dirac block $m_D$. The 
Majorana submatrix $M$ is composed of $\mu'$, $\mu$ and $\Lambda$: the $n'\times n'$, $n\times n$ and $n'\times n$ matrices 
respectively. Notice that $\epsilon$, $\mu_{ij}$ and $\mu'_{ij}$ are lepton-number-violating parameters.\footnote{This 
corresponds to the case in which $L\left(\nu_{\alpha L}\right)=L\left(N_{i=1,..,n}\right)=-L\left(N_{j=1,..,n'}\right)=1$. 
There are other possible lepton number assignments for $N_i$ and $N_j$, which are broken by different terms in the
Lagrangian. Our analysis is completely general and the neutrino mass matrix given in Eq.~(\ref{Mnu}) depends on all possible 
lepton-number-violating parameters.}
Another helpful, and widely used, basis is the one in which the Majorana submatrix for the sterile neutrinos is diagonal,
which we will denote with a tilde in the following discussion. In order to illustrate the relation 
between these two bases, let us consider the $n=n'=1$ case. Both bases are related trough the following rotation
which diagonalizes the Majorana submatrix $M$
\be
\label{rotation}
\tilde{M_\nu}=O M_\nu O^T = \left( \begin{array}{ccc} 0 &  \tilde{Y}_1^T v/\sqrt{2}& \tilde{Y}_2^T v/\sqrt{2} \\  \tilde{Y}_1 v/\sqrt{2} 
&  \tilde{M}_1 & 0 \\ \tilde{Y}_2 v/\sqrt{2}
 & 0 & \tilde{M}_2 
 \end{array} \right)\equiv \left( \begin{array}{cc} 0 &  \tilde{m}_D^T \\ \tilde{m}_D  
& \tilde{M} 
 \end{array} \right)\,,\;\;\;\;\;\;\;\; 
O= \left( \begin{array}{cc} 1 & 0  \\
0 &  A\\
\end{array} \right),
\ee
with $A$ being a $2\times 2$ orthogonal matrix with the rotation angle
\footnote{For simplicity, all the Majorana submatrix parameters 
in $M$ have been considered real.}
\be
\label{angle}
\tan\theta=\dfrac{\mu'-\mu + \sqrt{4\Lambda^2+(\mu'-\mu)^2 }}{2\Lambda}\,.
\ee
%
The Majorana masses $\tilde{M}_1$ and $\tilde{M}_2$ and the Yukawa couplings $\tilde{Y}_i$ are then given by
\bea
\label{Mgorrito}
\tilde{M}_{2,1}&=&\frac{1}{2}\left( \mu' + \mu \pm \sqrt{4\Lambda^2+(\mu'-\mu)^2 } \right),
\nonumber\\
\tilde{Y}_1&=&Y_1\cos\theta-\epsilon Y_2\sin\theta ~,
\nonumber\\
\tilde{Y}_2&=&Y_1\sin\theta+\epsilon Y_2\cos\theta ~. 
\eea
Of course, the analysis can be performed in any basis, but we will mainly work in the one in which the neutrino mass matrix 
is given by  Eq.~(\ref{Mnu}). 

Notice that the mass matrix given in Eq.~(\ref{Mnu}) is completely general. A particularly interesting set of models, 
included in Eq.~(\ref{Mnu}), are those studied and summarized in Ref.~\cite{Gavela:2009cd}, 
and which include the so-called \textit{inverse} or \textit{multiple} seesaw models
~\cite{Mohapatra:1986bd,Branco:1988ex,Shaposhnikov:2006nn,Kersten:2007vk}. The lepton number is assumed to be a good global 
symmetry only broken in the neutrino sector through the small lepton-number-violating terms $\mu$ and/or $\epsilon$. In these
models the light masses are ``naturally'' proportional to $\epsilon$ and/or $\mu$. 
Therefore, thanks to the suppression of the light neutrino masses coming from $\epsilon$ and $\mu$, the scale of NP 
given by $\Lambda$ can be lowered to the TeV level or even below. This allows sizable NP effects coming from the dimension-6 operator which, 
contrary to the dimension-5 one, does not present any extra suppression with $\epsilon$ and $\mu$, as it does not violate the lepton 
number~\cite{Gavela:2009cd}. 
Lepton-conserving processes very supressed in the SM as the rare decays 
are very promising channels to probe this kind of NP. Interesting recent analysis of the $\mu\rightarrow e \gamma$, 
$\mu\rightarrow eee$ and $\mu\rightarrow e$ conversions in the context of low-scale small lepton-number-violating models 
can be found in the literature~\cite{Blanchet:2009kk,Dinh:2012bp,Alonso:2012ji}. In Ref.~\cite{Ibarra:2011xn}, these sorts 
of models are studied in the context of the $0\nu\beta\beta$ decay using a different parametrization based on the Casas-Ibarra 
one~\cite{Casas:2001sr}, which parameterizes the neutrino mass matrix in terms of the light and heavy masses, the $U_{\alpha i}$ matrix 
and an orthogonal matrix $R$. Notice that the parametrization considered here is totally general and includes the Casas-Ibarra 
limit in which an approximate decoupling of light and heavy sectors is assumed.

Another interesting model included in Eq.~(\ref{Mnu}) is the so-called extended seesaw model~\cite{Kang:2006sn}. 
In these models $\mu'$ is the key parameter and is assumed to be larger than the rest of the parameters in Eq.~(\ref{Mnu}),
and more specifically, much larger than $\mu$ and $\epsilon Y_{2 \alpha} v$, defining the highest scale of the model. The term $\mu'$ introduces large 
lepton number violation which can help to achieve successful low-scale leptogenesis~\cite{Fukugita:1986hr} without the need of a
degenerate heavy neutrino spectrum~\cite{Kang:2006sn}. This large violation of the lepton number is not present in the inverse seesaw 
scenario, in which the lepton number is assumed to be a good approximate global symmetry.

In any case, we will not restrict our study to any particular value of the parameters or, in other words,
to any of the above mentioned specific limits. Nevertheless, for simplicity, 
we will consider the case in which only two fermion singlets ($n=n'=1$) are added. In any case, we expect that the general 
conclusions obtained in this work can be applied to models with larger number of right-handed neutrinos.

Finally, from neutrino oscillations, we know that it is not easy to accommodate the experimental data 
in the region of the parameter space between the limits: $\tilde{M_{i}}\gg \tilde{m}_D$ (\textit{seesaw limit})
and $\tilde{M_{i}}\ll \tilde{m}_D$ (\textit{pseudo-Dirac limit}). 
In fact, in Ref.~\cite{Donini:2011jh} it is shown how the constraints from neutrino oscillation 
experiments leave those limits as the only allowed regions for $n=n'=1$ and 
$\tilde{M_1}=\tilde{M_2}$. The region of the parameter space in between is ruled out and only the
pseudo-Dirac and seesaw limits survive. Reasonably 
extrapolating these results 
to the more general case with $\tilde{M_1}\neq\tilde{M_2}$ studied here, leaves the 
seesaw limit ($\tilde{M_i}\gg \tilde{m}_D$) as the only relevant part of the parameter space in the $0\nu\beta\beta$
decay context.\footnote{Of course, the Dirac limit will not be 
considered in this analysis where the $0\nu\beta\beta$ decay phenomenology is studied.} From now 
on, we will focus on the \textit{seesaw} limit. 
Notice, however, that this does not necessarily mean that $\tilde{M_i}$ have to be at the GUT or the TeV 
scale and can be considerably lighter~\cite{deGouvea:2005er,deGouvea:2006gz,Donini:2012tt}.

\section{Light neutrino masses and $0\nu\beta\beta$ decay}
\label{sec:masses}

For $\tilde{M_i}\gg \tilde{m}_D$, the light neutrino mass matrix is given 
at tree-level by
\be
\label{treemass}
m_{tree}\simeq -m_D^T M^{-1}m_D\simeq \frac{v^2}{2(\Lambda^2-\mu'\mu)}
\left(\mu Y_1^T Y_{1}+ \epsilon^2\mu' Y_2^T Y_{2}-\Lambda\epsilon (Y_2^T Y_{1}+Y_1^T Y_{2})\right)\,,
\ee
where $m_D$ and $M$ are the $2\times3$ Dirac and $2\times2$ Majorana submatrices, respectively, in Eq.~(\ref{Mnu}) for $n=n'=1$. 
Here, we have performed the standard ``seesaw" $m_D/M$ expansion, keeping the leading-order terms. We will discuss 
later if the higher-order corrections can be relevant. The contribution of the mostly active neutrinos to the 
$0\nu\beta\beta$ decay amplitude is proportional to the ``$ee$'' element of this effective mass matrix as
\bea
\label{light_exp}
A_{light}&\propto&\sum_{i=1}^{3} m_i U_{ei}^2 \mathcal{M}^{0\nu\beta\beta}(0) \approx -\left(m_D^T M^{-1}m_D\right)_{ee}  \mathcal{M}^{0\nu\beta\beta}(0)=
\nonumber\\
&=& \frac{ \mu Y_{1e}^2 + \epsilon Y_{2e}\left(\epsilon \mu' Y_{2e}-2\Lambda Y_{1e}\right)}{2(\Lambda^2-\mu'\mu)}\,v^2\mathcal{M}^{0\nu\beta\beta}(0)\,.
\eea
Therefore, the light neutrino contribution is strictly canceled as long as the parameters of the model satisfy the following 
relation:
\be
\mu Y_{1e}^2 + \epsilon Y_{2e}\left(\epsilon \mu' Y_{2e}-2\Lambda Y_{1e}\right)=0.
\label{secancela}
\ee
This condition is fulfilled for 
\be
\epsilon=\mu=0.
\label{condition}
\ee
Of course, it may also be satisfied for other choices of parameters, but 
$\epsilon=\mu=0$ is the most stable one under radiative corrections and higher-order terms in the expansion, as we 
will show later. From now on, we will assume that this cancellation condition is fulfilled. Setting $\epsilon$ and $\mu$ to 
zero also leads to vanishing tree-level active neutrino masses. However, light 
neutrino masses are expected to be generated at one loop if $\mu'$ is different from zero and breaks 
the lepton number, as we will see.

One could naively think that taking into account Eq.~(\ref{eq:GIMII}) would lead us to the same cancellation for the heavy 
neutrinos (see Eq.~(\ref{typeIamp})); however, the dependence of the NMEs on $m_I$
 avoids a complete cancellation, if the heavy neutrinos are not too degenerate. 

When the heavy neutrinos are above the $0\nu\beta\beta$ scale, $m_4,m_5\gg 100$ MeV, the heavy contribution to the $0\nu\beta\beta$ decay amplitude 
can be approximated as
\small
\bea
\label{heavyamp}
A_{extra}&\propto& \sum_I^{\rm extra} m_I U_{eI}^2 \mathcal{M}^{0\nu\beta\beta}(m_I)\propto -\left(m_D^T M^{-3}m_D\right)_{ee}
\\
&=& v^2\frac{\left(\mu^3+\Lambda^2(2\mu+\mu')\right)Y_{1e}^2-2\epsilon\Lambda\left(\Lambda^2+\mu'^2+\mu^2+\mu\mu'\right)Y_{1e}Y_{2e}+
\left( \mu'^2+\Lambda^2( \mu+2\mu' ) \right)\epsilon^2Y_{2e}^2 }{2 \left(\Lambda^2-\mu\mu'\right)^3},\nonumber
\eea
\normalsize
which reduces to

\be
\label{heavyamp2}
A_{extra}\propto \frac{v^2\mu'Y_{1e}^2 }{2 \Lambda^4}.
\ee
if the light neutrino contribution is canceled ($\epsilon=\mu=0$). Apparently, the above expression indicates that for large values of 
$\mu'$ and/or small enough $\Lambda$ the heavy neutrinos may give a relevant contribution to the $0\nu\beta\beta$ decay at tree 
level. At this point two interesting limits of Eq.~(\ref{Mnu}) arise:
\begin{itemize}
 \item  \textbf{Extended seesaw limit (\textit{ESS limit})}: $\mu'\gg \Lambda,\,m_D$. In view of Eq.~(\ref{heavyamp2}), this possibility appears quite 
 appealing. This limit is inspired by the so-called extended seesaw models and corresponds to a hierarchical spectrum for the heavy neutrinos:
\begin{eqnarray}
\begin{array}{ccccc} 
m_4\approx\tilde{M_1}\approx-\Lambda^2/\mu', & & & &U_{e4}\approx Y_{1e}v/\sqrt{2}\Lambda, \\ 
m_5\approx\tilde{M_2}\approx\mu', & & & & U_{e5}\approx Y_{1e}v/\sqrt{2}\mu',
\end{array}  
\end{eqnarray}
where we also show the corresponding mixing with the active neutrinos. In this regime, the lightest of the two heavy neutrinos dominates the heavy
contribution. Moreover, for large enough values of $\mu'$, $m_4$ becomes lighter than $100$ MeV, the NME takes its maximum value and the heavy 
contribution to the $0\nu\beta\beta$ decay becomes independent of $\Lambda$: 
\be
\label{Aheavyh}
A_{extra}\propto U_{e4}^2m_{4}\mathcal{M}^{0\nu\beta\beta}(0)\approx-\frac{Y_{1e}^2v^2}{2\mu'}\mathcal{M}^{0\nu\beta\beta}(0) ~.
\ee
\item 
\textbf{Inverse seesaw limit (\textit{ISS limit})}: $\Lambda\gg \mu',\,m_D$. This limit corresponds to one of the Minimal 
Flavor Violation (MFV) models studied in Ref.~\cite{Gavela:2009cd}. It is also related to the case analyzed in Ref.~\cite{Ibarra:2011xn},
where a different parameterization is used. In this
case the heavy neutrino spectrum is quasidegenerate, forming a quasi-Dirac pair:
\begin{eqnarray}
\begin{array}{ccccc} 
m_4\approx-m_5\approx\tilde{M_1}\approx -\tilde{M_2} \approx \Lambda, & & & & U_{e4} \approx U_{e5} \approx Y_{1e}v/2\Lambda,\\
\Delta \tilde{M}\equiv |\tilde{M_2}|-|\tilde{M_1}|\approx \mu', & & & & 
\end{array}  
\end{eqnarray}
and we can expect lepton-number-violating processes such as neutrinoless double beta decay 
to be controlled by $\mu^\prime$.
\end{itemize}

If all the heavy neutrinos are located below the  $0\nu\beta\beta$ scale, a cancellation driven by Eq.~(\ref{eq:GIMII}) is expected at tree-level, 
as we have already mentioned. This cancellation applies in general as long as all the heavy neutrinos are in the light regime, 
including the two limits distinguished above. 

The approximation made in Eq.~(\ref{heavyamp}), $\mathcal{M}^{0\nu\beta\beta}(m_I)\propto 1/m_I^2$, does not apply 
if one of the heavy neutrinos (or both) is lighter than (or close to) $\sim 100$ MeV. However, as we have already commented, we will not 
restrict the analysis to any particular value of the sterile neutrino masses. This is the reason why we have made use of a numerical 
computation for the NME in which no approximation for the neutrino mass dependence has been considered. Notice, for instance, that 
the phenomenology for heavy masses around $100$ MeV can be very interesting and the approximation 
$\mathcal{M}^{0\nu\beta\beta}(m_I)\propto 1/m_I^2$ is not very accurate in that region.

In summary, at tree-level the light neutrino masses are independent of $\mu'$ (and $\Lambda$) for $\epsilon=\mu=0$, 
being actually zero. However, lepton-number-violating processes such as $0\nu\beta\beta$ decay are sensitive to these parameters
 and $\mu'$ in particular. The idea behind Refs.~\cite{Ibarra:2011xn,Mitra:2011qr} is to exploit this apparent decoupling 
between the heavy and light contributions in order to have a measurable effect in the  $0\nu\beta\beta$ decay coming from the heavy 
side. In the following, we will check if a heavy dominant contribution is really possible once the relevant 
corrections and experimental constraints are taken into account.

\section{Higher-order corrections in the seesaw expansion}

Only the leading-order in $m_D/M$ has been considered in the expansion performed in Eq.~(\ref{treemass}). We now check if 
the higher-order corrections may induce any relevant effects once the tree-level cancellation for the light masses takes place. 
The next-to-leading-order contributions to the light neutrino masses
can be written as~\cite{Grimus:2000vj}:

\be
\delta m=\frac{1}{2}m_{tree}\, m_D^\dagger M^{-2} m_D + \frac{1}{2}\left(m_{tree}\, m_D^\dagger M^{-2} m_D\right)^T,
\ee
where $m_{tree}$ is the leading-order contribution given by $m_{tree}=-m_D^T M^{-1} m_D$. 
As they are proportional to the leading-order active neutrino mass $m_{tree}$, they are completely 
irrelevant for $\mu=\epsilon=0$. In fact, the light neutrino masses vanish for $\mu=\epsilon=0$ at all orders
in the expansion~\cite{Grimus:2000vj,Adhikari:2010yt}. Contrary to the $\mu=\epsilon=0$ case, other choices of the parameters which 
satisfy the cancellation condition given in Eq.~(\ref{secancela}) are flavor dependent, giving as a result nonvanishing 
higher-order corrections. 

On the other hand, the factor $m_D^\dagger M^{-2} m_D/2$ is nothing but the coefficient of the effective $d=6$ operator 
obtained when the heavy neutrinos are integrated out of the theory~\cite{Broncano:2002rw}. This coefficient,  
which induces deviations from the unitarity of the $3\times 3$ lepton mixing matrix, is independent of 
$\mu'$ when the light neutrino cancellation ($\mu=\epsilon=0$) takes place. Therefore, for $\mu=\epsilon=0$, the  $d=6$ operator does not introduce any relevant $\mu'$-dependent 
deviation from unitarity, and $\mu'$ can escape from the corresponding constraints~\cite{Antusch:2006vwa,Antusch:2008tz}, 
even if $\mu'\gg\Lambda$. 

\section{One-loop corrections}

The one-loop corrections can be of two different types: renormalizable (i.e. the running of the parameters) 
or finite. In this section we will study both, starting with the renormalizable corrections. The analysis will be done after 
electroweak symmetry breaking (EWSB).

\subsection{Renormalizable one-loop corrections}
 
We are mainly interested in the running behavior of the parameters $\mu$ and $\epsilon$, since they drive the light neutrino mass cancellation, and 
$\mu'$ and $\Lambda$ which are the key parameters associated with the heavy contribution. We have performed the computation in the 
basis in which the neutrino mass matrix takes the form given by Eq.~(\ref{Mnu}), in such a way that the one-loop running equations~\cite{Pirogov:1998tj,Haba:1999ca,Casas:1999tp}
for these parameters can be directly obtained:

\begin{eqnarray}
Q \dfrac{d \left( \epsilon Y_{2\alpha } \right)}{dQ}&=&\frac{\epsilon}{(4\pi)^2}
\left[\left( T-\frac{9}{4}g^2-\frac{3}{4}g^{'2} \right) Y_{2\alpha}
-\frac{3}{2}\, Y_{2\beta}\left((Y_l^{\dagger}Y_l)_{\beta\alpha}- Y^{*}_{1\beta}Y_{1\alpha}\right)
+\frac{3}{2}\epsilon^{2}\,Y_{2\beta}Y^{*}_{2\beta}Y_{2\alpha} \right],
\nonumber\\
Q\dfrac{d \mu}{dQ}&=& \frac{2\epsilon}{(4\pi)^2}\left[\Lambda\,Y^{*}_{1\beta}Y_{2\beta}+\mu\epsilon\,Y^{*}_{2\beta}Y_{2\beta}\right],
\nonumber\\
Q\dfrac{d \mu'}{dQ}&=& \frac{2}{(4\pi)^2}\left[\mu'\,Y^{*}_{1\beta}Y_{1\beta}+\epsilon\Lambda\,Y^{*}_{2\beta}Y_{1\beta}\right],
\nonumber\\
Q\dfrac{d \Lambda}{dQ}&=& \frac{1}{(4\pi)^2}\left[\Lambda\, Y^{*}_{1\beta}Y_{1\beta}+\epsilon\left(
\mu'\, Y^{*}_{1\beta}Y_{2\beta}+\mu \,Y^{*}_{2\beta}Y_{1\beta}+ \Lambda\,\epsilon\,Y^{*}_{2\beta}Y_{2\beta} \right) \right],
\label{running2}
\end{eqnarray}
where $T=\text{Tr}\left(3Y_u^{\dagger}Y_u+3Y_d^{\dagger}Y_d+Y_l^{\dagger}Y_l+Y^{\dagger}Y\right)$ and
$g$ and $g^\prime$ are the $SU(2)_L$ and $U(1)_Y$ gauge coupling constants of the SM. We do not need to solve 
the equations to realize that the effect of the one-loop renormalizable corrections to $\mu$ and $\epsilon$ is suppressed by the tree-level 
values of $\epsilon$ or $\mu$. This means that the cancellation of the light active neutrino masses is stable under one-loop renormalizable 
corrections, as expected, as a Majorana mass coupling for the active neutrinos is not allowed at tree-level.
For vanishing $\epsilon$ and $\mu$ at tree-level, the light neutrino masses keep being zero independently of the running 
of the parameters (even for huge tree-level inputs of $\mu'$). This is no longer true once the finite corrections are taken into account, 
as we show in the next subsection. 

\subsection{Finite one-loop corrections}
\label{finito}

Indeed, after EWSB, a Majorana mass for the active neutrinos is generated through finite one-loop corrections. 
Of course, the other Yukawa and Majorana couplings among the active 
and sterile neutrinos also get finite corrections, but their contribution to the light neutrino masses vanishes for 
$\mu=\epsilon=0$. This contribution is proportional to the finite one-loop 
corrections to $\mu$ and $\epsilon Y_{2\alpha }$ (see Eq.~(\ref{treemass})). Since the sterile neutrinos 
only couple to the Higgs, via the Yukawas, the one-loop corrections to $\mu$ (the Majorana coupling between 
$\overline{N_{2}} N_{2}^{c}$) and the Yukawa couplings between $N_2$ and $\nu_{\alpha L}$ ($\epsilon Y_{2\alpha }$) are 
proportional to $\epsilon Y_{2\beta }$ and vanish in the limit $\mu=\epsilon=0$. Therefore, the dominant contribution to 
the light neutrino masses comes from the Majorana mass generated for the active neutrinos and is given 
by~\cite{Pilaftsis:1991ug,Grimus:2002nk,AristizabalSierra:2011mn} 
\begin{equation}
\label{finitos}
\left(\delta m_{LL}\right)_{\alpha\beta}= 
\frac{1}{(4 \pi v)^2}\left(\tilde{m}_D^T\right)_{\alpha i} \tilde{M}_i
\left\{\frac{3\ln\left(\tilde{M}_i^2/M^2_Z\right)}{\tilde{M}_i^2/M^2_Z-1}+\frac{\ln\left(\tilde{M}_i^2/M^2_H\right)}{\tilde{M}_i^2
/M^2_H-1}\right\}\left(\tilde{m}_D\right)_{i\beta}\,,
\end{equation}
where $\tilde{m}_D$ and $\tilde{M}=\text{diag}\left(\tilde{M}_1,\tilde{M}_2\right)$ are the Dirac and Majorana submatrices 
respectively, written in the basis in which the Majorana submatrix is diagonal, $M_Z$ is the mass of the $Z$ boson, and $M_H$ is the Higgs boson mass. 
Notice that no expansion has been performed in order to obtain this result. The structure of the correction is similar to the 
tree-level masses but in this case no cancellation takes place for $\mu=\epsilon=0$.  

\begin{figure}[h!]
\includegraphics[width=0.6\textwidth,angle=0]{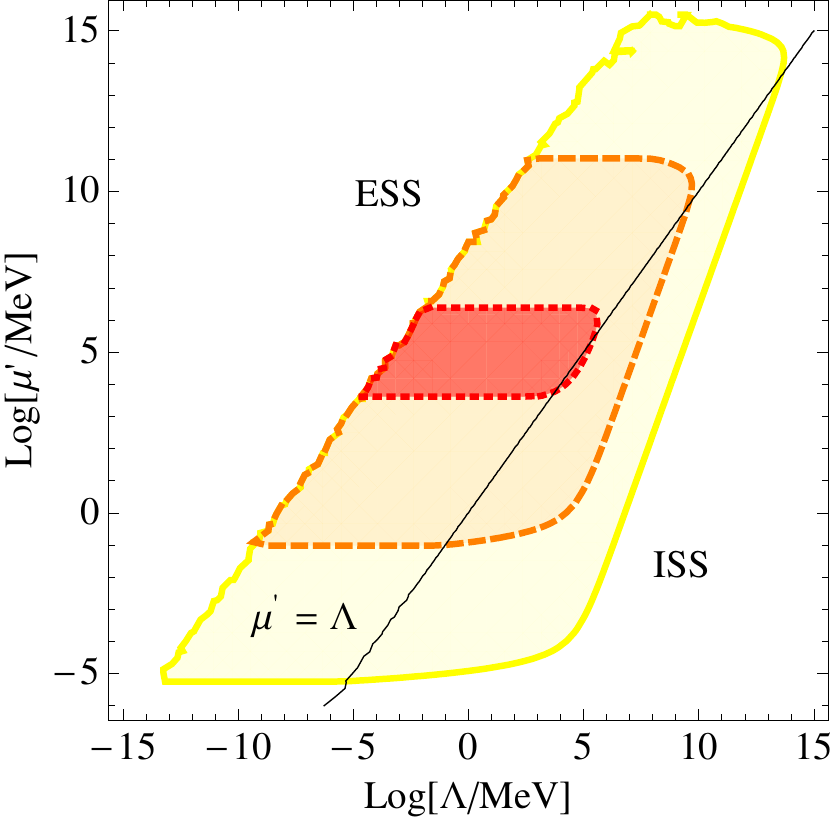} 
\caption{\label{mLL} The colored region in the $\mu^\prime$-$\Lambda$ plane corresponds to $|\delta m_{LL}\left(\mu=\epsilon=0\right)|>0.1$ eV.  The yellow area (solid edges), orange (dashed),
and red (dotted) stands for
$Y_{1\alpha}=10^{-1}$, $Y_{1\alpha}=10^{-3}$, and $Y_{1\alpha}=10^{-5}$, respectively.}
\end{figure}

In particular, Eq.~(\ref{finitos}) can be conveniently written in 
the $\mu=\epsilon=0$ limit as
\bea
\label{delta_finite}
\delta m_{LL}
&=&\frac{1}{(4 \pi)^2}\frac{Y_1^T Y_1}{2} \left\{
\left( \frac{3\tilde{M_1}\ln\left(\tilde{M_1}^2/M^2_Z\right)}{\tilde{M_1}^2/M^2_Z-1}+\frac{\tilde{M_1}\ln\left(\tilde{M_1}^2/M^2_H\right)}
{\tilde{M_1}^2/M^2_H-1}\right)\cos^2\theta\right.
\nonumber\\
&+&
\left.\left(\frac{3\tilde{M}_2\ln\left(\tilde{M}_2^2/M^2_Z\right)}{\tilde{M}_2^2/M^2_Z-1}+\frac{\tilde{M}_2\ln\left(\tilde{M}_2^2/M^2_H\right)}
{\tilde{M}_2^2/M^2_H-1}\right)\sin^2\theta \right\}.
\eea
where $\tilde{M}_{2,1}$ are the eigenvalues of the Majorana mass term given by Eq.~(\ref{Mgorrito}), and $\theta$ is the rotation 
angle given by Eq.~(\ref{angle}), both evaluated for $\epsilon=\mu=0$. In Fig.~\ref{mLL}, we show the 
region of the parameter space $\mu^\prime$-$\Lambda$ given by $|\delta m_{LL}\left(\mu=\epsilon=0\right)|>0.1$ eV for different values of the Yukawa couplings.
In order to understand better the implications of Eq.~(\ref{delta_finite}), we have obtained approximate expressions for two relevant limits:
\begin{itemize}
\item $\Lambda \gg \mu',M_H,M_Z$. We have
\be
\label{mLL_ISS}
\delta m_{LL}\approx \frac{1}{(4 \pi)^2}\,\frac{Y_1^T Y_1}{2}\,\frac{M_H^2+3M_Z^2}{\Lambda^2}\,\,\mu'\,.
\ee
As we have already discussed in Sec.~\ref{sec:masses}, this case is included in the ISS limit and corresponds to a MFV model in which $\mu$, $\epsilon$ and
$\mu'$ are lepton-number-violating parameters. What we observe here is that although the tree-level light neutrino masses 
cancel for $\epsilon=\mu=0$, they are generated at one loop and are proportional to the only 
lepton-number-violating parameter different from zero, $\mu^\prime$, as expected, since the neutrino masses also
violate this symmetry.
\item $\mu' \gg \Lambda\gg M_H,M_Z$. In this case, one finds
\be
\label{mLL_ESS}
\delta m_{LL}\approx \frac{1}{(4 \pi)^2}\frac{Y_1^T Y_1}{2}\left[\frac{3M_Z^2}{\mu'}\ln\left(\frac{\Lambda^4}{M_Z^4}\right)+
\frac{M_H^2}{\mu'}\ln\left(\frac{\Lambda^4}{M_H^4}\right)\right].
\ee
This case is included in the ESS limit discussed in Sec.~\ref{sec:masses}. Here, the one-loop light neutrino masses depend mildly on $\Lambda$ 
and are suppressed by $\mu'$. Again, this can be understood in terms of a lepton symmetry: $\mu'$ is suppressing the violation of the lepton number 
at low energies in such a way that in the limit $\mu'\rightarrow\infty$, the symmetry is completely restored in the effective theory.
\end{itemize}

In the next section, we will study the phenomenological consequences of Eq.~(\ref{delta_finite}) in the context of the 
$0\nu\beta\beta$ decay without considering any expansion on the parameters. It is important to remark here that once the tree-level cancellation takes place, only one mass
is generated at one-loop, and at least two light masses are necessary to explain the light neutrino spectrum obtained in 
neutrino oscillation experiments. This is easy to solve: simply adding another fermion singlet to the model would allow us to generate 
the necessary extra light mass. However, for simplicity, we will keep studying the simpler case with only two extra sterile neutrinos.

\section{New Physics dominant contribution to $0\nu\beta\beta$ decay and one-loop neutrino masses}

Once the relevant one-loop corrections are taken into account, the Lagrangian is modified
to
\begin{eqnarray}
\mathscr{L} &=& \mathscr{L}_\mathrm{SM}+ \mathscr{L}_\mathrm{kin}-\frac{1}{2} 
\overline{N_{i}} M_{ij} N_{j}^{c}-\frac{1}{2}(\delta m_{LL})_{\alpha\beta}
\overline{\nu_{\alpha L}}\nu_{\beta L}^c
 -y_{i\alpha}\overline{N_{i}} 
\widetilde \phi^\dagger
L_\alpha +\text{h.c}.\,.
\end{eqnarray}
Consequently, Eq.~(\ref{eq:GIMII}), which comes from the diagonalization of the neutrino mass matrix,
is also modified to the following one-loop version:
\begin{equation}
\sum_i^{\rm light} m_i U_{ei}^2 + \sum_I^{\rm extra} m_I U_{eI}^2 = (\delta m_{LL})_{ee} .
\label{eq:GIM1loop}
\end{equation}
%
Notice that here $U$ diagonalizes the neutrino mass matrix including the one-loop
corrections. In the case of interest, when $\epsilon=\mu=0$, the light neutrino masses associated with the mostly active neutrinos are 
determined by $\delta m_{LL}$. The tree-level condition $\sum_I^{\rm extra} m_I U_{eI}^2=0$ remains true at the one-loop 
level but, as discussed in Sec.~\ref{sec:masses}, heavy neutrinos could have a sizable effect on $0\nu\beta\beta$ decay thanks to 
the NME dependence on the heavy masses. However, at one loop, the NP contribution to the $0\nu\beta\beta$ decay and 
the light neutrino masses are related, as they depend on the same parameters, in particular $\mu'$, $\Lambda$ and 
$Y_{1\alpha}$. Their decoupling achieved at tree-level does not remain true once radiative corrections are included. 
Consequently, the heavy parameters cannot be chosen arbitrarily as to dominate $0\nu\beta\beta$ decay but 
are constrained by light neutrino masses, which also contribute to the process. 

In principle, the radiative corrections dependent on the transferred momentum $p$ also have to be considered. This corrections
are of two types: (i) proportional to $p$ or $p^3$; (ii) dependent on $p^2$. The first come from the $W$ and charged Goldstone boson 
corrections to the neutrino propagator and vanish by chirality. The 
second are associated with the $Z$ and Higgs boson corrections to the propagator and are negligible in the region of heavy masses 
under consideration.

\begin{figure}[h!]
\includegraphics[width=0.6\textwidth,angle=0]{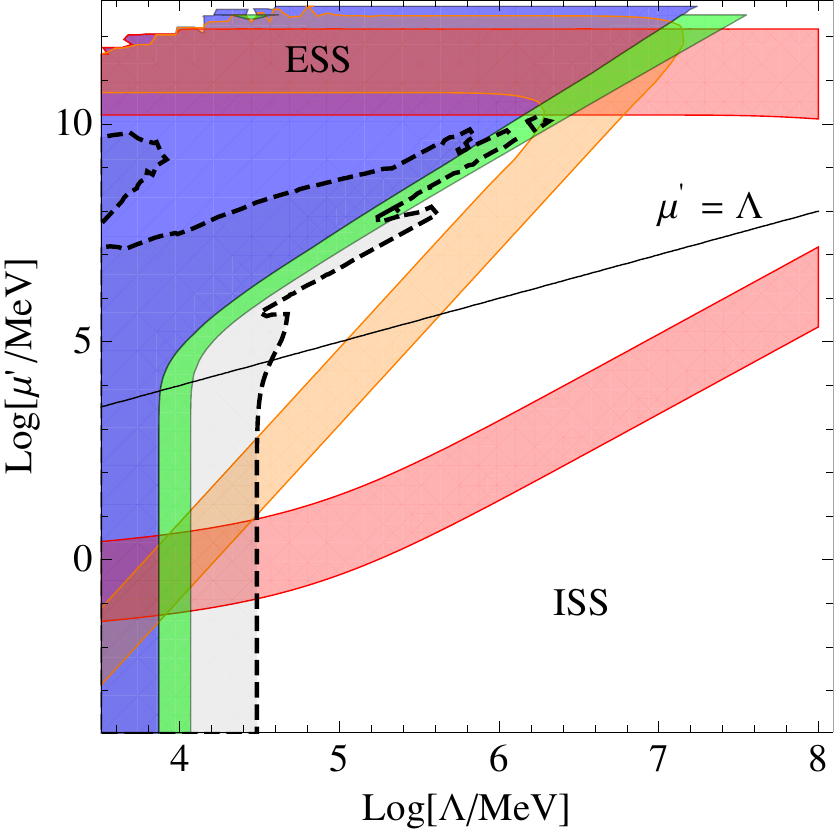} 
\caption{\label{combinacion} Impact of one-loop corrections on $0\nu\beta\beta$ decay for $Y_{1 \alpha }=10^{-3}$. The red band is the 95\%C.L. 
allowed region for the one-loop generated light neutrino masses bounded by cosmology and neutrino oscillations. 
The orange band is the 95\%C.L. region of the parameter space in which the heavy neutrino contribution is between the present bound from 
EXO and the sensitivity of the next-to-next generation of $0\nu\beta\beta$ experiments. Blue (green) stands for
the region in which the ratio $r$ between the heavy and light contributions is $r>5$ ($1<r<5$). The grey region inside the dashed 
black line is the parameter space ruled out at the 95\% C.L. by the constraints on the mixing.}
\end{figure}

In the rest of the section, we will study under which conditions it may (or may not) be possible to have a dominant heavy neutrino 
contribution. We will pay special attention to the impact of the one-loop corrections and the 
experimental constraints on the parameters of the model. To illuminate the interplay among all these 
factors, we will first analyze the particular case of $Y_{1\alpha}=10^{-3}$ showing our results in 
Fig.~\ref{combinacion}. 

First of all, we are assuming that the model under consideration provides the dominant source of light 
neutrino masses. In principle, they should be in agreement with neutrino oscillations data 
but, since we are generating at one loop only one light neutrino mass we only impose a conservative lower bound 
on the nontrivial eigenvalue of Eq.~(\ref{delta_finite}) given by the solar splitting, $\sqrt{\Delta m^2_{sol}}$. Moreover, the 
absolute neutrino mass scale experiments impose an upper bound on the same combination of parameters,
and as reference value we take the 95\%C.L. upper bound on the light neutrino masses
from cosmology~\cite{Komatsu:2010fb}, $m_\nu=0.58$ eV. 
Since we are analyzing the case in which the tree-level active neutrino masses cancel ($\epsilon=\mu=0$), these bounds can be directly 
translated into bounds 
on $\mu'$ and $\Lambda$ as a function of $Y_{1 \alpha }$. They are shown in Fig.~\ref{combinacion}
as the red band. The Higgs mass, $m_H$, has been fixed to $125$ GeV in all the calculations, 
as suggested by the recent LHC results~\cite{:2012gu,:2012gk}. Notice that if no lower bound is imposed, as would be the 
case if the light neutrino masses came from some other mechanism, the outer region of the red band would not be excluded. 
Our conclusions remain valid also in this case, as we will discuss later. The constraint on the light neutrino masses shown in Fig.~\ref{combinacion} 
can be understood by analytically taking into account 
the approximate expressions derived in the previous section. In the ISS limit, 
$\sqrt{\Delta m^2_{sol}}<\delta m_{LL}< 0.58$ eV scales as $\mu'/\Lambda^2$ in agreement with 
Eq.~(\ref{mLL_ISS}). For $\mu'\gg \Lambda$, in the ESS limit, it becomes mainly 
independent of $\Lambda$ according to Eq.~(\ref{mLL_ESS}).

The heavy neutrino contribution to $0\nu\beta\beta$ decay, given by \;$A_{heavy}\propto\sum_{I=4,5}U_{eI}^2m_{I}\mathcal{M}^{0\nu\beta\beta}(m_I)$, can be computed 
by diagonalizing the mass matrix in Eq.~(\ref{Mnu})\footnote{Notice that the corrections on the heavy 
mixing $U_{eI}$ due to the one-loop effects are negligible, since $\tilde{M}\gg \delta m_{LL},\tilde{m}_D$.} and using the NME data calculated as
a function of the neutrino masses~\cite{nme}. The diagonalization can be easily performed in 
the $\epsilon=\mu=0$ limit. This contribution has to respect the present experimental 
bound and, in order to be phenomenologically interesting, should be within the reach of future 
$0\nu\beta\beta$ decay experiments such as CUORE~\cite{CUORE}, EXO~\cite{Auger:2012ar}, GERDA~\cite{GERDA}, 
KamLAND-Zen~\cite{KamLANDZen:2012aa}, MAJORANA~\cite{Gaitskell:2003zr}, NEXT~\cite{Diaz:2009zzb} or 
Super-NEMO~\cite{Barabash:2004pp}. This constraint is shown 
as the orange band in Fig.~\ref{combinacion}: the 95\%C.L. region of the parameter space 
in which the heavy neutrino contribution 
is between the present bound from EXO~\cite{Auger:2012ar} 
($0\nu\beta\beta$ decay in $^{136}$Xe), which using the corresponding shell model NME is $|m_{\beta\beta}|< 0.53$ eV, 
and the future (optimistic) sensitivity of the next-to-next generation 
of $0\nu\beta\beta$ decay experiments, taken to be $m_{\beta\beta}=10^{-2}$ eV. 
The shape of the heavy contribution contour can also be easily understood from the discussion in Sec.~\ref{sec:masses}. 
The heavy contribution scales as $\mu^\prime/\Lambda^4$, following Eq.~(\ref{heavyamp2}) closely 
until, for $\mu'\gg\Lambda$, it becomes independent of $\Lambda$ in agreement with Eq.~(\ref{Aheavyh}). In the ISS region, both heavy neutrinos 
have masses larger than the  $0\nu\beta\beta$ scale, and Eq.~(\ref{heavyamp2}) holds. As expected from comparing 
Eq.~(\ref{heavyamp2}) and Eq.~(\ref{mLL_ISS}), in this region the slope of the heavy contribution contour is twice the 
$\sqrt{\Delta m^2_{sol}}<\delta m_{LL}< 0.58$ eV one. For $\mu'\gg\Lambda$, we enter the ESS limit and eventually the lightest of the two heavy 
masses becomes lighter than the $0\nu\beta\beta$ scale ($\sim 100$ MeV), while the heaviest one is too heavy to give a relevant contribution 
to the process. In this region, the ``heavy'' contribution is thus dominated by the sterile neutrino lighter than $100$ MeV,  for which the 
corresponding NME takes the maximum value, and is independent of $\Lambda$ (see Eq.~(\ref{Aheavyh})). This dominant behavior of the sterile 
neutrino lighter than $100$ MeV will be confirmed later in Fig.~\ref{intersection}, as we will 
explain below.

Figure~\ref{combinacion} also highlights the region of the parameter space for which the ratio $r$ 
between the heavy and mostly active contribution to $0\nu\beta\beta$ decay, defined as $r\equiv|A_{heavy}/A_{light}|$,  
is between 1 and 5 (green region) or larger than 5 
(blue region). 
The active contribution is determined by the one-loop correction to the 
light neutrino masses: $A_{active}\propto \left(\delta m_{LL}\right)_{ee}\mathcal{M}^{0\nu\beta\beta}(0)$. 
From Eqs.~(\ref{heavyamp2}), (\ref{Aheavyh}) and (\ref{delta_finite}), it is clear that $r\equiv|A_{heavy}/A_{light}|$ should be 
basically independent of the Yukawa couplings. 

\begin{figure}[h!]
\begin{tabular}{ccc}
  \includegraphics[width=0.33\textwidth,angle=0]{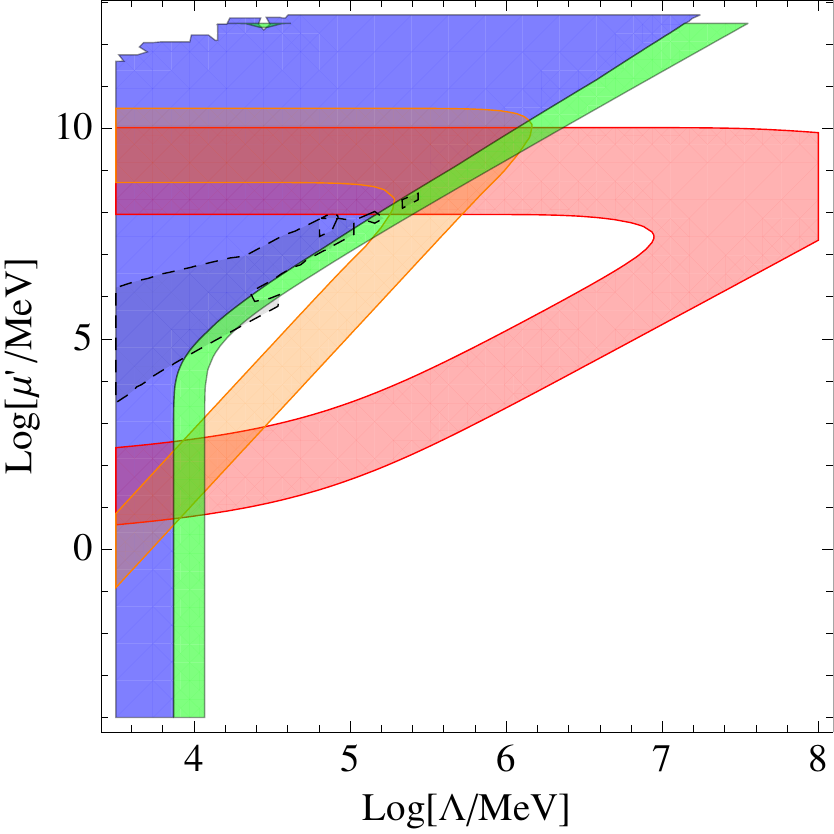} &
    \includegraphics[width=0.33\textwidth,angle=0]{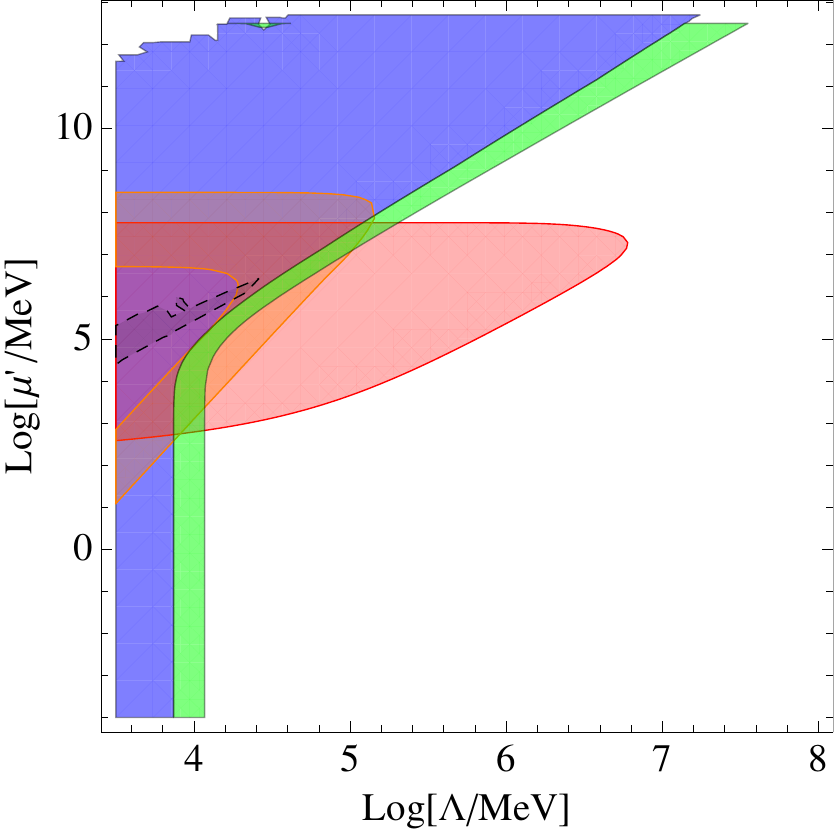} &
      \includegraphics[width=0.33\textwidth,angle=0]{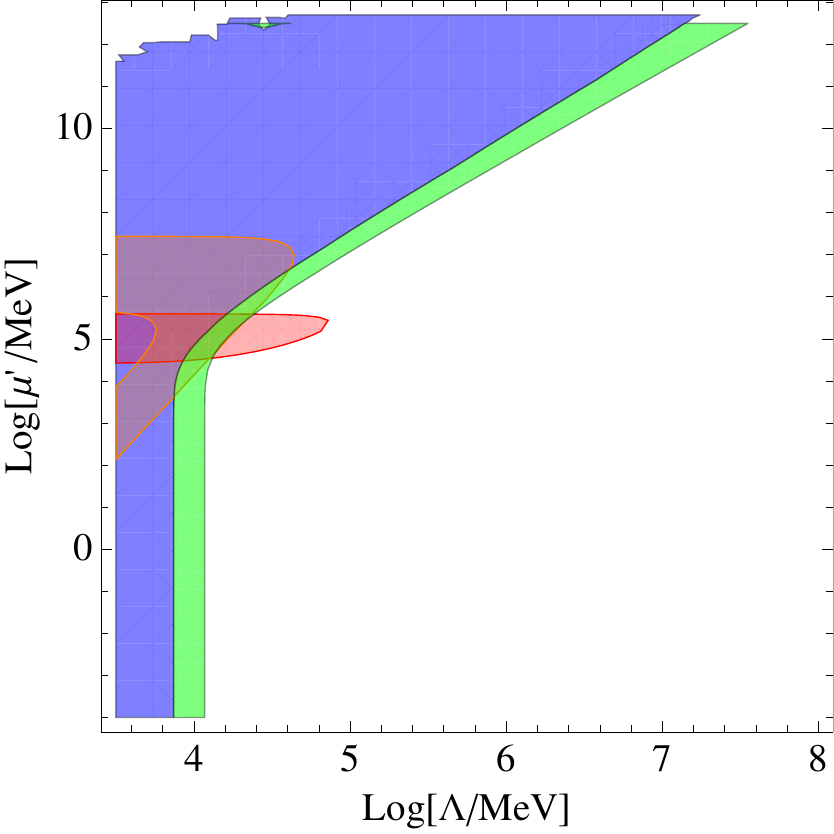}
 \end{tabular}\caption{\label{fig2} Same as in Fig.~\ref{combinacion}, for  $Y_{1 \alpha }=10^{-4}$ (left),  $Y_{1 \alpha }=10^{-5}$ (center), 
 and $Y_{1 \alpha }=3\cdot 10^{-6}$ (right).}
\end{figure}

Finally, the information coming from the experiments that constrain the mixing between the active and heavy 
neutrinos is also included in Fig.~\ref{combinacion}. The grey region inside the dashed line is excluded at the 95\% C.L. 
by the constraints on the mixing extracted from weak decays (summarized in~\cite{Atre:2009rg}) and 
non-unitarity bounds~\cite{Antusch:2006vwa,Antusch:2008tz}.

As shown in Fig.~\ref{combinacion}, it is possible to have a dominant and measurable contribution from the heavy neutrinos to $0\nu\beta\beta$ 
decay, keeping the light neutrino masses under control. Ignoring for the sake of discussion the constraints on the heavy mixing, 
this takes place in the two intersections among the red, the orange, and the blue regions, which lie in two interesting 
limits already discussed in Secs.~\ref{sec:model} and \ref{sec:masses}:
\begin{itemize}
\item i) \textbf{ISS limit}: $\Lambda\gg \mu',Y_{1 \alpha } v$.  
The heavy neutrinos are quasidegenerate, 
and their contribution to the process is proportional to the splitting, given by $\mu'$. 
Once the constraints on the mixing,
$U_{e4} \sim U_{e5} \sim Y_{1 e } v/ 2 \Lambda$, are properly taken into account, the ISS limit is ruled out.
\item ii) \textbf{ESS limit}: $\mu' \gg \Lambda, Y_{1 \alpha } v$.
In this case, the lightest of the extra neutrinos has a mass lower than the neutrinoless double beta decay
exchange momentum and dominates the process.
\end{itemize}

 \begin{figure}[h!]
  \includegraphics[width=0.6\textwidth,angle=0]{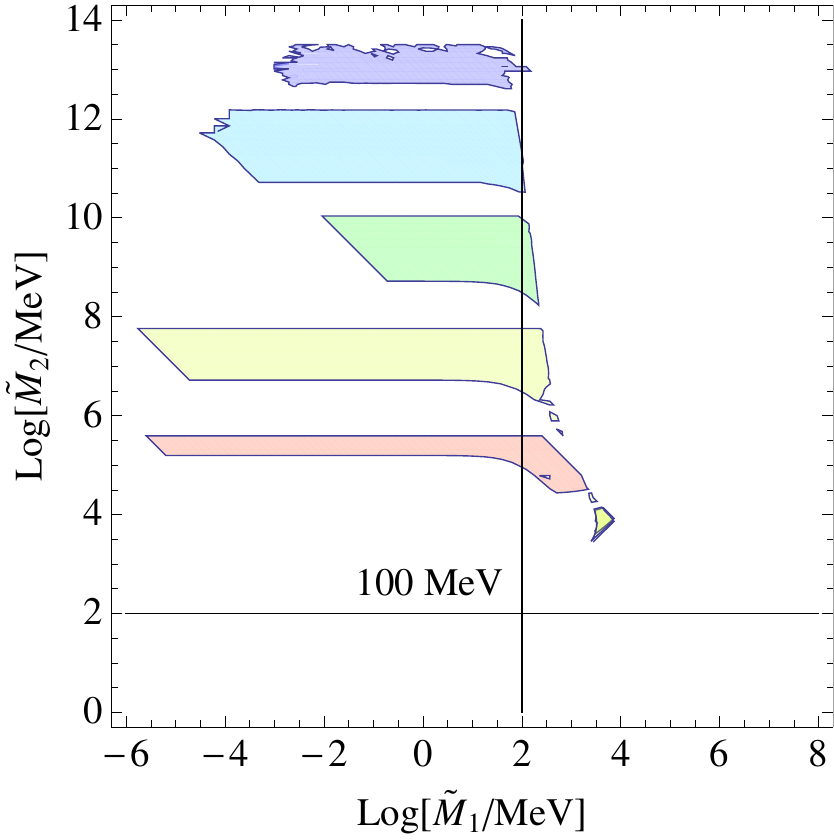} 
\caption{\label{intersection} Region of the parameter space, $\tilde{M_2}\approx m_5$ vs $\tilde{M_1}\approx m_4$, in which a dominant and measurable 
contribution of the heavy neutrinos is feasible, respecting bounds from neutrino oscillations, absolute neutrino 
mass scale experiments and weak decays. From top to bottom, the blue, cyan,
green, yellow and red areas stand for $Y_{1\alpha}=10^{-2}$, $10^{-3}$, $10^{-4}$, $10^{-5}$, and $3\cdot10^{-6}$, respectively. The black 
lines correspond to $\tilde{M_1}=100$ MeV and $\tilde{M_2}=100$ MeV.}
\end{figure}
 
We have chosen $10^{-3}$ as the input value of $Y_{1\alpha}$ in Fig.~\ref{combinacion} as an 
example that allowed us to illuminate the discussion. The results for $Y_{1\alpha}=10^{-2}-10^{-3}$ are 
similar, but we have checked that for values 
of the Yukawa couplings larger than $10^{-2}$, a dominant contribution from the heavy neutrinos is not 
possible and can be at most of the same order as the contribution from light neutrinos. In Fig.~\ref{fig2}, we show the plots analogous 
to Fig.~\ref{combinacion} but for smaller values of the Yukawa couplings: $10^{-4}$ (left), $10^{-5}$ (center), and 
$3\cdot10^{-6}$ (right). For these values, the heavy neutrino mixing
is small enough to satisfy the bounds coming from weak decays. We observe that 
the ratio between the light and heavy contributions is independent of the Yukawa couplings
as expected. However, each of them separately depends strongly on that input. The region of the parameter space in 
which we have a measurable heavy contribution decreases with the Yukawa coupling, as is also the case
for the red region, in which the light neutrino masses keep being under control. We have checked
that between $10^{-6}$ and $10^{-8}$, a dominant and measurable contribution of the heavy neutrinos may still be possible, but the 
light neutrino masses generated at one-loop are smaller than $\sqrt{\Delta m_{sol}^2}$. For values of the Yukawa couplings smaller than 
$10^{-8}$, the heavy contribution is too suppressed to be experimentally accessible.

The information given in Figs.~\ref{combinacion} and \ref{fig2} is summarized in Fig.~\ref{intersection}, where we show the region of the parameter 
space in which a dominant and measurable contribution of the heavy neutrinos is possible, respecting at the same time the bounds on heavy mixing from 
weak decays and non-unitarity~\cite{Atre:2009rg,Antusch:2006vwa,Antusch:2008tz},
and keeping light neutrino masses in the region between $\sqrt{\Delta m_{solar}^2}$ and their upper 
bound extracted from Ref.~\cite{Komatsu:2010fb}.
 Although the tree-level 
cancellation for the light neutrino masses is taking place, once the one-loop corrections are taken into account, a dominant contribution from the heavy 
neutrinos cannot occur for larger or smaller values of the Yukawa couplings than the ones shown in Fig.~\ref{intersection}. Notice that 
this dominant contribution is mainly possible only in the hierarchical seesaw scenario mentioned above 
($|\tilde{M_1}| \lesssim 100$ MeV $\ll |\tilde{M_2}|$), where the lightest sterile neutrino gets a mass smaller than (or around) 
$100$ MeV and dominates the process. Indeed, this result is not surprising: in Ref.~\cite{Blennow:2010th} it was shown that, in the case in which
the cancellation of the light neutrino contribution does not occur, a hierarchical heavy spectrum like this is necessary in order to have a relevant contribution from 
the heavy neutrinos at tree-level. We have checked in this work that this conclusion, obtained at tree-level, can be extended to
the case in which a cancellation of the light contribution takes place at tree-level if the one-loop level corrections are included in the analysis. Nevertheless, there is an exception to these conclusions. For $Y_{1\alpha}\approx10^{-4}-10^{-5}$, there is still a 
tiny region in which the heavy contribution could dominate when the heavy neutrinos are quasidegenerate and around $5$ GeV 
(ISS region).

A comment is in order: the qualitative conclusions just depicted above are not affected significantly if the lower bound on the 
one-loop light neutrino masses imposed here ($\delta m_{LL}>\sqrt{\Delta m^2_{sol}}$) is not assumed in the 
analysis. In such a case, the allowed regions in Fig.~\ref{intersection} become a bit larger 
(vertically), and a dominant heavy neutrino contribution would be still possible for 
$Y_{1\alpha}=10^{-6}-10^{-8}$. Also in this case,
a hierarchical spectrum with $|\tilde{M_1}| \lesssim 100$ MeV $\ll |\tilde{M_2}|$ is required, with the possible exception
of having a quasidegenerate spectrum with $|\tilde{M_1}|\sim|\tilde{M_2}|\sim 5$ GeV (in the same tiny region of the parameter
space). This can be easily understood from Figs.~\ref{combinacion}-\ref{fig2}: eliminating the lower bound on $\delta m_{LL}$ would 
mean that the outer region of the red bands would not be forbidden any more.

Finally, it should be remarked that the results presented in this section are not modified 
if a different upper bound on the light neutrino masses from the one used in our 
analysis ($m_\nu = 0.58$ eV~\cite{Komatsu:2010fb}) is considered. Modifying this upper bound would be reflected in a 
slight modification of the inner boundary of the red bands in Figs.~\ref{combinacion}-\ref{fig2}, which have a marginal impact 
on the final results. (The intersection among the different contours is not affected.) In particular, we have checked that
the conclusions drawn here remain valid if an upper bound from cosmology of $m_\nu = 0.36$ eV~\cite{dePutter:2012sh} is 
considered instead in the analysis. 

\section{Summary and Conclusions}

The possibility of having a dominant contribution from heavy neutrinos to $0\nu\beta\beta$ 
decay, when a cancellation of the tree-level light neutrino contribution takes place, 
has recently received much attention~\cite{Ibarra:2011xn,Mitra:2011qr}. In this work we 
have carefully analyzed this possibility in the general framework of type-I seesaw models. We have considered a general 
parameterization of the neutrino mass matrix 
which allowed us to explore the whole parameter space, identifying particularly interesting limits
such as the inverse and extended seesaw models. 
 We have shown which conditions have to be satisfied for a stable cancellation of the tree-level light neutrino 
contribution, allowing the heavy neutrinos to dominate the process at tree-level.
 We have studied the relevant 
corrections that may arise in this context. The finite one-loop corrections to the light neutrino masses turn out to be
very relevant. Although logarithmic,
their contribution to the $0\nu\beta\beta$ decay rate tends to dominate very easily.
We have found that the heavy neutrinos can give the main contribution to the 
process only for a very hierarchical heavy neutrino
spectrum with masses below and above the  $0\nu\beta\beta$ scale $\sim100$~MeV, which would match an extended seesaw like model.
The ``heavy'' neutrino contribution is in fact completely dominated by the lightest sterile neutrinos with mass $\lesssim 100$ MeV, which is not 
suppressed by the NME. This result coincides with the general conclusions of the tree-level analysis performed when no 
cancellation takes place~\cite{Blennow:2010th}. Quantitatively, we have obtained that values 
of the Yukawa couplings between $10^{-2}$ and $10^{-6}$ ($10^{-8}$) are necessary, if a lower bound on the one-loop neutrino masses of
$\sqrt{\delta m^2_{sol}}$ (no lower bound) is imposed in the analysis. 
We qualitatively agree with part of the conclusions drawn in Ref.~\cite{Mitra:2011qr}: 
the extended seesaw scenario might accommodate a 
relevant ``heavy'' neutrino contribution. 
Nevertheless, our general conclusions clarify an important detail: in Ref.~\cite{Mitra:2011qr} it was hypothesized that this 
may happen with all the heavy neutrinos above the $0\nu\beta\beta$ decay scale while we conclude that the heavy 
spectrum needs to contain states in both regimes, below (or close to) and above 100 MeV.

An interesting exception arises for quasidegenerate heavy neutrinos with masses around $5$ GeV which may give the dominant contribution in a tiny region of 
the parameter space for Yukawa couplings in the range $10^{-4}-10^{-5}$. 
In agreement with Refs.~\cite{Ibarra:2011xn,Mitra:2011qr}, we confirm that a relevant contribution to the $0\nu\beta\beta$ decay may come from 
a seesaw scenario with a quasidegenerate heavy neutrino spectrum, which corresponds to an inverse seesaw like model. However, we also 
show that this possibility is rather unlikely, since it can only take place in a very particular and small region of the parameter space. 

Our results can be understood from the point of view of lepton number conservation. Even if the light neutrino 
contribution cancels out at tree-level, in order to have a measurable heavy contribution an important violation of the lepton 
number should be introduced through the heavy sector. This violation of the lepton number may not be reflected in the tree 
level light neutrino masses but appears naturally at the one-loop level, making more difficult a dominant heavy contribution.

Finally, we should remark that 
our analysis was performed considering two fermion singlets, and only one light neutrino mass was generated at 
one loop. In order to generate the two light neutrino masses required to explain neutrino oscillations, one has two 
options: (i) not considering a complete cancellation for the tree-level light neutrino masses, being only partial 
($\mu$ and/or $\epsilon$ small parameters but different from zero), or (ii) adding more fermion singlets to the model. 
In case (i), tree-level light neutrino masses are generated. They might be very small, but the seesaw constraint given by 
Eq.~(\ref{eq:GIM1loop}) would leave the light active neutrinos as the dominant mechanism in $0\nu\beta\beta$ decay, at least 
if a fine-tuned cancellation between the tree-level and one-loop neutrino contribution is not invoked. Again, in this scenario
a dominant contribution from the heavy neutrinos can be expected mainly for $|\tilde{M_1}| \lesssim 100$ MeV $\ll |\tilde{M_2}|$
. In case (ii), adding an even number of sterile neutrinos with the opposite lepton number ($n=n'$) would allow a complete 
cancellation for the light tree-level neutrino masses, generating at the same time two or more light neutrino masses at 
one loop. This kind of model was studied recently in Ref.~\cite{Dev:2012sg}, where six sterile neutrinos were considered 
($n=n'=3$). These models match the ISS limit studied here, but obviously the
number of free parameters is larger ($\mu'$ and $\Lambda$ are $n'\times n'$ matrices, and $Y_1$ is a $n'\times 3$ matrix). 
The lepton number violation, required to have a relevant heavy neutrino contribution, generates at the same time 
light neutrino masses at one loop and the latter will typically dominate in $0\nu\beta\beta$ decay, independently from 
the number of generations considered. Relevant exceptions are the cases highlighted in our study or possible further 
fine-tuned cancellations among the contributions due to different generations. Therefore, we expect similar results 
to the ones presented in this work, although a detailed analysis beyond the scope 
of this work would be necessary in order to take into account the just-mentioned fine-tuned cancellations and show the 
constraints in the corresponding larger parameter space.

\section*{Note}
During the completion of this work an analysis which considered the generation of \,neutrino 
masses at the loop-level in inverse seesaw models was presented in Ref.~\cite{Dev:2012sg}.

\begin{acknowledgments}
We thank R. Alonso, E. Fern\'andez-Mart\'inez, B. Gavela, P. Hern\'andez, J. Men\'endez, and J. Yepes for useful discussions. 
We are also grateful to S. Petcov for useful discussions and important remarks.
We would like to thank the Galileo Galilei Institute for Theoretical Physics, and SP also thanks SISSA for the hospitality during the
completion of this work. This work was partially supported by the European projects EURONU (CE212372), EuCARD (European Coordination 
for Accelerator Research and Development, Grant Agreement number 227579) and the ITN INVISIBLES (Marie Curie Actions, PITN- GA-2011-289442).
\end{acknowledgments}

\end{document}